\documentclass[fleqn,10pt]{wlscirep}

\def\be{\begin{equation}}
\def\ee{\end{equation}}
\def\ba{\begin{array}}
\def\ea{\end{array}}

\title{Optimal Universal Uncertainty Relations}

\author[1,*]{Tao Li}
\author[2,3,**]{Yunlong Xiao}
\author[4]{Teng Ma}
\author[5,3]{Shao-Ming Fei}
\author[6,2,3]{Naihuan Jing}
\author[3]{Xianqing Li-Jost}
\author[5]{Zhi-Xi Wang}
\affil[1]{School of Science, Beijing Technology and Business University, Beijing 100048, China}
\affil[2]{School of Mathematics, South China University of Technology, Guangzhou, Guangdong 510640, China}
\affil[3]{Max Planck Institute for Mathematics in the Sciences, Leipzig 04103, Germany}
\affil[4]{State Key Laboratory of Low-Dimensional Quantum Physics and Department of Physics, Tsinghua University, Beijing 100084, China}
\affil[5]{School of Mathematical Sciences, Capital Normal University, Beijing 100048,  China}
\affil[6]{Department of Mathematics, North Carolina State University, Raleigh, NC 27695, USA}
\affil[*]{corresponding author: lt881122@sina.com}
\affil[**]{corresponding author: mathxiao123@gmail.com}

\begin{abstract}
We study universal uncertainty relations and present a method called joint probability distribution diagram to improve the
majorization bounds constructed independently in [Phys. Rev. Lett. \textbf{111}, 230401 (2013)] and [J. Phys. A. \textbf{46}, 272002 (2013)].
The results give rise to state independent uncertainty relations satisfied by any nonnegative Schur-concave functions. On the other hand,
a remarkable recent result of entropic uncertainty relation is the direct-sum majorization relation.
In this paper, we illustrate our bounds by showing how they
provide a complement to that in [Phys. Rev. A. \textbf{89}, 052115 (2014)].
\end{abstract}
\begin{document}
\flushbottom
\maketitle
\thispagestyle{empty}

\section*{Introduction}

Uncertainty relations \cite{Heisenberg1} are of profound significance in quantum mechanics and quantum information theory.
Various important applications of uncertainty relations have been discovered such as entanglement detection \cite{Guhne2}, steering inequalities \cite{Schneeloch4} and quantum cryptography \cite{Koashi5,Renes6,Tomamichel7}. The well-known form of the Heisenberg's uncertainty relations, given by Robertson \cite{Robertson8}, says that the standard deviations of the observables $\Delta A$ and $\Delta B$ satisfy the following inequality,
\be\label{1}
\Delta A\cdot \Delta B \geq |\langle \psi |[A,B]|\psi\rangle|/2.
\ee
As a consequence of the uncertainty relations, it is impossible to determine the exact values of the two incompatible observables simultaneously.
However, the lower bound in the above uncertainty inequality may become trivial if the measured state $|\psi\rangle$ belongs to the nullspace of the commutator $[A,B]$.

In fact, the uncertainty relations provide a limitation on how much information one can obtain by measuring a physical system, and
can be characterized in terms of the probability distributions of the measurement outcomes.
In order to overcome the drawback in the product form of variance base uncertainty relations, Deutsch \cite{Deutsch9} introduced the entropic uncertainty relations,
which were later improved by Maassen and Uffink \cite{Maassen10}: $H(A)+H(B)\geq -2\log c(A,B)$, where $H$ is the Shannon entropy,
$c(A,B)=\max \limits_{m,n}|\langle a_m|b_n\rangle|$ is maximum overlap between the basis elements $\{|a_m\rangle\}$ and $\{|b_n\rangle\}$ of the eigenbases
of $A$ and $B$, respectively. Recently, the Maassen-Uffink bound has been surprisingly improved by Coles and Piani \cite{CP11}, Rudnicki, Pucha{\l}a and \.{Z}yczkowski \cite{RPZ12}, for a review on entopic uncertainty relations see references \cite{Wehner13,CP14}.

Friedland, Gheorghiu and Gour \cite{Friedland15} proposed a new concept called ``universal uncertainty relations'' which are not limited to considering only the well-known entropic functions such as Shannon entropy, Renyi entropy and Tsallis entropy, but also any nonnegative Schur-concave functions. On the other hand, Pucha{\l}a, Rudnicki and \.{Z}yczkowski \cite{RPZ16} independently used majorization technique to establish  entropic uncertainty relations similar to ``universal uncertainty relations''. Let ${|a_m\rangle}^d_{m=1}$ and ${|b_n\rangle}^d_{n=1}$ be orthonormal bases of a $d$-dimensional Hilbert space $H$. Denote by $p_m(\rho)=\langle a_m|\rho|a_m\rangle$ and $q_n(\rho)=\langle b_n|\rho|b_n\rangle$ the probability distributions obtained by measuring the state $\rho$ with respect to these bases, which constitute
two probability vectors ${\bf p}(\rho)=(p_1,p_2,...,p_d)$ and ${\bf q}(\rho)=(q_1,q_2,...,q_d)$, respectively.
It has been shown that the tensor product of the two probability vectors ${\bf p}(\rho)$ and ${\bf q}(\rho)$
is majored by a vector ${\bf \omega}$ independent from the  state $\rho$,
\begin{equation}\label{e:UUR}
{\bf p}(\rho)\otimes {\bf q}(\rho)\prec {\bf \omega}, ~~ for ~~ any ~~ \rho,
\end{equation}
where ``$\prec$" stands for ``majorization'': ${\bf x}\prec {\bf y}$ in $R^d$ if
$\sum \limits^k \limits_{j=1} x_k^{\downarrow}\leq \sum \limits^k \limits_{j=1} y_k^{\downarrow}$ for all $1\leq k\leq d-1$, and
$\sum \limits^d \limits_{j=1} x_k^{\downarrow}= \sum \limits^d \limits_{j=1} y_k^{\downarrow}$.
The down-arrow vector $x^{\downarrow}$ denotes that the components of ${\bf x}$ are rearranged in descending order,
$x_{1}^{\downarrow}\geq x_{2}^{\downarrow}\geq\cdots \geq x_{d}^{\downarrow}$.
The $d^2$-dimensional vector ${\bf \omega}$ is given by
\begin{equation}\label{e:omega1}
\omega = (\Omega_1, \Omega_2-\Omega_1, \ldots, \Omega_d-\Omega_{d-1}, 0, \ldots, 0),
\end{equation}
where
\begin{equation}\label{e:omega2}
\Omega_k=\max\limits_{I_k}\max\limits_{\rho}\sum\limits_{(m,n)\in I_{k}}p_m(\rho)\,q_n(\rho)
\end{equation}
with $I_k \subset [d]\times [d]$ being a subset of $k$ distinct pairs of indices $(m, n)$ and $[d]$ is the set of the natural numbers from $1$ to $d$.
The outer maximum is over all subsets $I_k$ with cardinality $k$ and the inner maximum runs over all density matrices.

Eq. (\ref{e:UUR}) is called a universal uncertainty relation, as for any uncertainty
measure $\Phi$, a nonnegative Schur-concave function, one has that
\be\label{phi}
\Phi[p(\rho)\otimes q(\rho)]\geq \Phi(\omega),~~ for ~~ any ~~ \rho.
\ee
The universal uncertainty relation (UUR) (\ref{e:UUR}) generates infinitely many uncertainty relations, for each $\Phi$,
in which the right hand side provides a single lower bound.

In relation (\ref{e:UUR}), the state independent vector $\omega$ decided by $\Omega_k$  in Eq. (\ref{e:omega2}) is too hard to evaluate explicitly in general,
as it is involved with a highly nontrivial optimization problem. For this reason, only an approximation $\tilde{\Omega}_k$ of $\Omega_k$ has been presented \cite{Friedland15,RPZ16} to construct a weaker majorization vector $\tilde{\omega}$. Naturally, how to find a stronger approximation than previous works becomes an interesting open question.

\section*{Results}

We first introduce a scheme called ``joint probability distribution diagram'' (JPDD) to consider the optimization problem involved in calculating $\Omega_k$. Next, we present a stronger approximation by proposing an analytical formula for $\Omega_k$. To facilitate presentation,
%And in order to read facilitated,
we denote our stronger approximation as $\Omega_k$ without ambiguity. %In special case, if necessarily, we will manifest for the reader.
All uncertainty relations considered in the paper will be
in the absence of quantum side information.

To construct the joint probability distribution diagram, we associate each summand $p_i(\rho)q_j(\rho)$ in $\Omega_k$ to a box located at the position
$(i, j)$. Then the summation in $\Omega_k$ corresponds to certain region of boxes (or rather lattice points)
in the first quadrant. We configure the region in a combinatorial way. Suppose that
$p_1\geq p_2\geq \cdots \geq p_d$, $q_1\geq q_2 \geq \cdots \geq q_d$. Consider the following $d\times d$-matrix
\begin{equation}\label{pq}
\left(
\begin{array}{cccc}
  p_1q_1 & p_1q_2 & \cdots & p_1q_d \\
  p_2q_1 & p_2q_2 & \cdots & p_2q_d \\
  \vdots & \vdots & \ddots & \vdots \\
  p_dq_1 & p_dq_2 & \cdots & p_dq_d
\end{array}
\right),
\end{equation}
where the entries descend along the rows and columns by assumption. Now, we
${\rm use\ a\ box\ } {\renewcommand{\arraystretch}{0.96} \begin{tabular}{|c|}
                               \hline
                               \ \\
                               \hline
                             \end{tabular}}
{\rm\ to}$ represent an entry of the matrix.
A {\it shadow} or {\it grey box} in the JPDD means the corresponding entry in the matrix. For example,
the top left shadow of the block box specifies the entry $p_1q_1$, see Fig. 1. Thus the region corresponding to the summation in $\widetilde{\Omega}_k$ will be
a special region of the rectangular matrix.

Our scheme, JPDD, provides a combinatorial method to compute the special region with respect to the $\Omega_k$. First, it is easy to see that the top upper
left box in JPDD is the maximal element, i.e. $\Omega_1$,
since $p_1q_1-p_iq_j\geq p_iq_1-p_iq_j=p_i(q_1-q_j)\geq 0$.
The main idea is that each exact solution of $\Omega_k$ corresponds to a particular
region in this matrix.

Suppose that the $k$-th region is found, i.e. $\Omega_k$ is obtained, then the next $(k+1)$-th region
is obtained from the $k$-th region by adding a special box, which must be ``connected'' with
certain boundary of the $k$-th region. This iterative procedure enables us to
compute all $\Omega_k$. Before proving the statement rigorously, we first introduce some terminologies.

\begin{figure}[ht]
\centering
\label{e:omega0}
\includegraphics[width=0.12\textwidth]{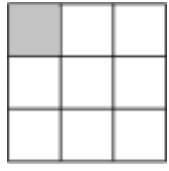}
\caption{{The left-top shadow of the block box specifies the entry $p_1q_1$.}}
\end{figure}

[{\sf Definition 1}] (Different boxes)~~
Two boxes (matrix elements) $p_iq_j$ and $p_kq_l$ are said to be {\it different} if they
occupy different positions in JPDD, namely, $i\neq k$ or $j\neq l$. The Fig. 2 shows three examples of different boxes. Note that it may happen that even if the numerical values of $p_iq_j$ and $p_kq_l$ are the same, but graphically
they are treated as different boxes. ``different'' and ``same'' do not imply their quantitative relation.
For example, $p_1q_1$ may equal to $p_1q_3$ in general.

\begin{figure}[ht]
\centering
\includegraphics[width=0.36\textwidth]{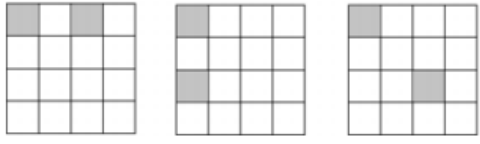}
\caption{{Different boxes.}}
\end{figure}

\begin{figure}[ht]
\centering
\includegraphics[width=0.26\textwidth]{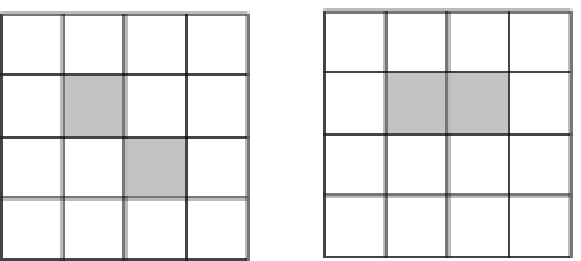}
\caption{{Connectedness.}}
\end{figure}

\begin{figure}[ht]
  \centering
  \label{fig:1}
\includegraphics[width=4.5cm]{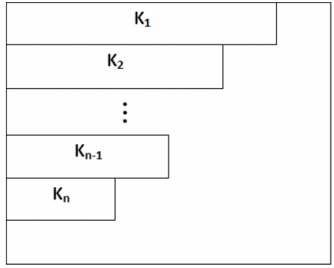}
\caption{{$\Omega_{k_1, k_2,\cdots , k_n}^d$ related joint probability distribution diagram.}}
\end{figure}

[{\sf Definition 2}] (Connectedness).~~ Two boxes $p_iq_j$ and $p_kq_l$ in JPDD are {\it connected} if there
does not exist any box $p_mq_n$, different from both $p_iq_j$ and $p_kq_l$, such that
$\min \{p_iq_j, p_kq_l\}<p_mq_n< \max \{p_iq_j, p_kq_l\}$. For example, different boxes $p_iq_j$ and $p_kq_l$ are connected if $p_iq_j=p_kq_l$.
For $d=4$, %we give an example to explain this definition.
if $p_1>p_2>p_3>p_4$, $q_1>q_2>q_3>q_4$, $p_iq_j \neq p_kq_l$, $p_mq_n \neq p_iq_j$, $p_mq_n \neq p_kq_l$,
for any $i,j,k,l\in \{1,2,3,4\}$, then $p_2q_2$ and $p_3q_3$ are not connected while $p_2q_3$ and $p_2q_2$ are connected, see Fig. 3.

[{\sf Definition 3}] (Connected region).~~ A set of different boxes is called a {\it region}, denoted by $\mathfrak{A}$.
A region $\mathfrak{A}$ is {\it connected} if $\forall\, p_iq_j\in\max \mathfrak{A}$, then either $p_{i-1}q_j \in \mathfrak{A}$ or $p_iq_{j-1} \in \mathfrak{A}$,
where $\max \mathfrak{A}$ is the maximal value of all the elements in $\mathfrak{A}$. Note that
the region of boxes corresponding to a $\Omega_k$ must contain the top-left element
$p_1q_1$ in JPDD as its largest element.

For any probability vector ${\bf p}(\rho)$ on a $d$-dimensional Hilbert space with $p_i=\langle a_i|\rho|a_i\rangle$,
let $p_1^{\prime}=\max\limits_{i}\{p_i\}$, $p_{2}^{\prime}=\max\limits_{\begin{subarray}{c}p_i\neq p_j^{\prime}\\j=1\end{subarray}}\{p_i\}$, $\cdots$,
$p_{k}^{\prime}=\max\limits_{\begin{subarray}{c}p_i\neq p_j^{\prime}\\j=1,2\cdots k-1\end{subarray}}\{p_i\}$.
Similarly, $q_1^{\prime}, \cdots, q_{k}^{\prime}$ are defined similarly for another probability vector
${\bf q}(\rho)$ on the same Hilbert space.
For any sequence $k_1, \cdots, k_n$, $1\leq n \leq d$, we define
$
\Omega_{k_1, k_2,\cdots , k_n}^d=p_1^{\prime}(q_1^{\prime}+q_2^{\prime}+\cdots q_{k_1}^{\prime})+p_2^{\prime}(q_1^{\prime}+q_2^{\prime}+\cdots q_{k_2}^{\prime})
+\cdots +p_n^{\prime}(q_1^{\prime}+q_2^{\prime}+\cdots q_{k_n}^{\prime}).
$
In particular, if $p_1\geq \cdots \geq p_d$, $q_1\geq\cdots \geq q_d$, then $\Omega_{k_1, k_2,\cdots , k_n}^d=p_1(q_1+q_2+\cdots q_{k_1})+p_2(q_1+q_2+\cdots q_{k_2})+\cdots +p_n(q_1+q_2+\cdots q_{k_n})$, which can be configured by the Fig. 4. In a JPDD when the first $k$ boxes are chosen, the next (maximal) $(k+1)$-th box must appear at the top left corner in the unoccupied region, we give this as follow lemma:\\
\begin{figure}[ht]
  \centering
  \label{fig:1}
\includegraphics[width=7.6cm]{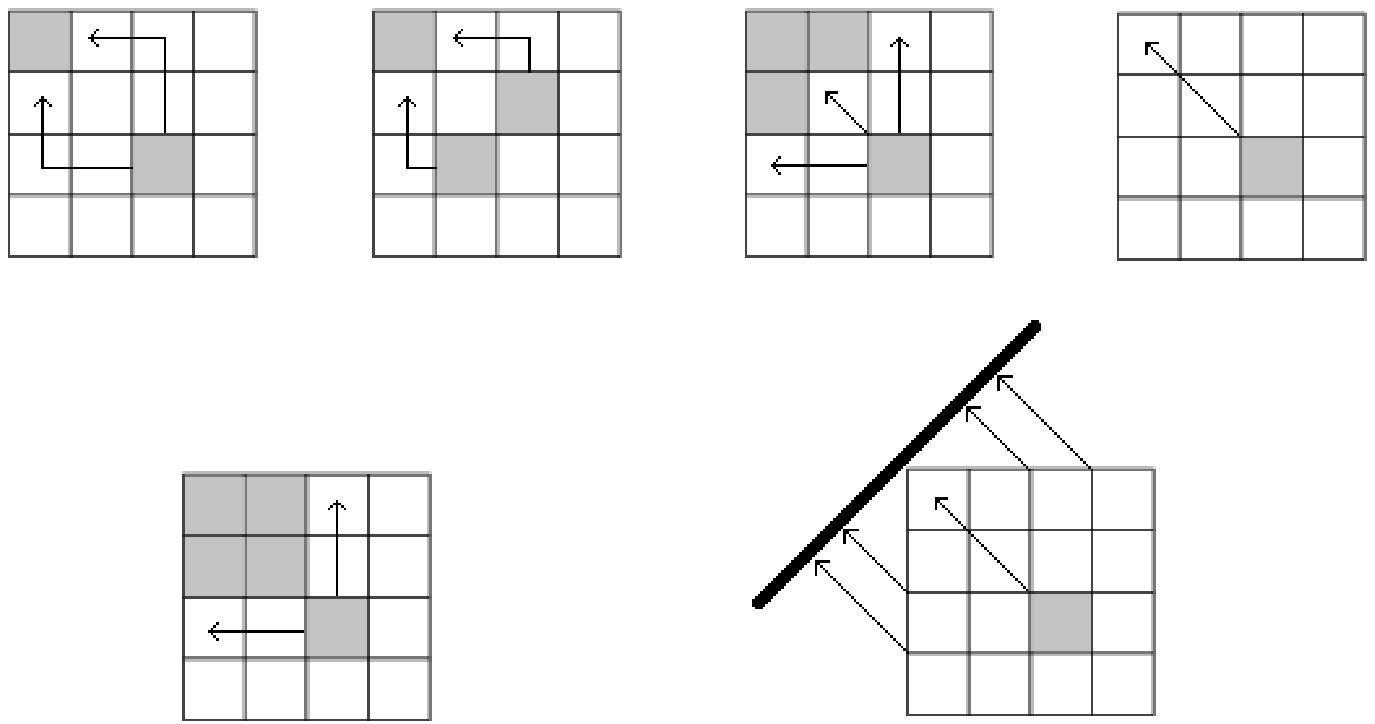}
\caption{The (k+1)-th box is fixed by $\Omega_k$. The arrows show the position of the (k+1)-th box.}
\end{figure}

\begin{figure}[ht]
\centering
\includegraphics[width=0.48\textwidth]{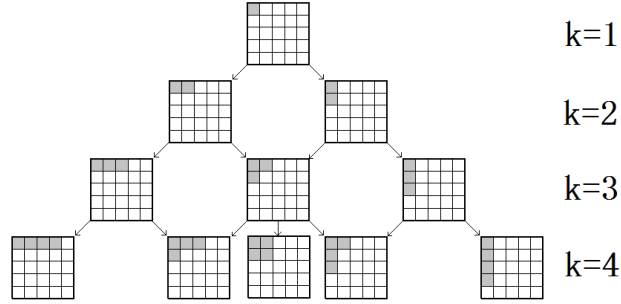}
\caption{{The tree diagrams to get the JPDDs in the $k$-th row by adding one shadow box to the JPDDs in the $k-1$ row.}}
\end{figure}
[{\sf Lemma}]: The maximal $k$ boxes for $\Omega_k$ in the JPDD can be selected to form a connected region. ~~~~~~~~~~~~~~~~~~~~~~~~~~~~~~~~~~~~~~$\blacksquare$
\\

\noindent Lemma gives a way to get $\Omega_{k+1}$ from $\Omega_k$ in a JPDD.
As an example, we show how to get $\Omega_3$ from $\Omega_2$.
Set $c_{11}=\max\limits_{m,n}|\langle a_m|b_n\rangle|$, $c_{21}=\max\limits_{m,n^{'}\neq n}\sqrt{|\langle a_m|b_n\rangle|^2+|\langle a_m|b_{n^{'}}\rangle|^2}$,
$c_{22}=\max\limits_{n,m^{'}\neq m}\sqrt{|\langle a_m|b_n\rangle|^2+|\langle a_{m^{'}}|b_{n}\rangle|^2}$.
If $c_{21}\geq c_{22}$, %$c_{21}-c_{22}\geq 0$,
then $\Omega_3=\max\{\Omega_3^d,\\ \Omega_{2,1}^d\}$. If
$c_{21}\leq c_{22}$, then $\Omega_3=\max\{\Omega_{1,1,1}^d, \Omega_{2,1}^d\}$. That is, if $\Omega_2=\Omega_2^d$, then $\Omega_3=\max\{\Omega_3^d,\Omega_{2,1}^d\}$. If $\Omega_2=\Omega_{1,1}^d$, then $\Omega_3=\max\{\Omega_{1,1,1}^d,\Omega_{2,1}^d\}$. Namely,
$\Omega_3=\max\{\Omega_{3}^d,\Omega_{2,1}^d,\Omega^d_{1,1,1}\}$. Thus $\Omega_3$ is determined by $\Omega_2$.

\noindent In general, $\Omega_{k-1}=\Omega^d_{k_1,k_2,\cdots ,k_n}$, subjecting to $\sum\limits^{n}\limits_{i=1}k_i=k-1$. By Lemma it follows that
\begin{equation}
\Omega_k=\max\{p_1q_{k_1+1},p_2q_{k_2+1},\cdots,p_nq_{k_n+1},p_{n+1}q_1\}+\Omega_{k-1}
=\max\{\Omega^d_{k_1+1,k_2,\cdots,k_n},\Omega^d_{k_1,k_2+1,\cdots,k_n},\cdots,\Omega^d_{k_1,k_2,\cdots,k_n+1}\},
\end{equation}
which gives an iterative formula of $\Omega_{k}$ in terms of $\Omega_k^d$'s.

%\begin{align*}
%\Omega_k&=\max\{p_1q_{k_1+1},p_2q_{k_2+1},\cdots,p_nq_{k_n+1},p_{n+1}q_1\}+\Omega_{k-1}\\
%&=\max\{\Omega^d_{k_1+1,k_2,\cdots,k_n},\Omega^d_{k_1,k_2+1,\cdots,k_n},\cdots,\Omega^d_{k_1,k_2,\cdots,k_n+1}\},
%\end{align*}

We list in Figs. 5,6 all the possible $\Omega_k$ for $k=1,2,..., 4$. The above example to get $\Omega_3$ from $\Omega_2$ corresponds to move from the second row to the third. Now we are ready to show the main result.\\

{\sf [Theorem]} The quantities $\Omega_{k}$ are given by

\begin{equation}\label{e:omega3}
\begin{aligned}
\Omega_{k} &=\frac{1}{4}\max\limits_{\begin{subarray}{c}\sum \limits_{i=1}^n k_i=k\\k_1\geq ...\geq k_n\end{subarray}}
[(\max\limits_{\begin{subarray}{c}|R|=n\\|S|=k_n\end{subarray}}||\sum \limits_{m\in R}|a_m\rangle \langle a_m|+\sum \limits_{n\in S}|b_n\rangle \langle b_n|||^2_{\infty}
-\max\limits_{\begin{subarray}{c}|R|=n-1\\|S|=k_n\end{subarray}}||\sum \limits_{m\in R}|a_m\rangle \langle a_m|+\sum \limits_{n\in S}|b_n\rangle \langle b_n|||^2_{\infty})\\[2mm]
&+(\max\limits_{\begin{subarray}{c}|R|=n-1\\|S|=k_{n-1}\end{subarray}}||\sum \limits_{m\in R}|a_m\rangle \langle a_m|+\sum \limits_{n\in S}|b_n\rangle \langle b_n|||^2_{\infty}
-\max\limits_{\begin{subarray}{c}|R|=n-2\\|S|=k_{n-1}\end{subarray}}||\sum \limits_{m\in R}|a_m\rangle \langle a_m|+\sum \limits_{n\in S}|b_n\rangle \langle b_n|||^2_{\infty})\\[2mm]
&+\cdots+(\max\limits_{\begin{subarray}{c}|R|=2\\|S|=k_{2}\end{subarray}}||\sum \limits_{m\in R}|a_m\rangle \langle a_m|+\sum \limits_{n\in S}|b_n\rangle \langle b_n|||^2_{\infty}
-\max\limits_{\begin{subarray}{c}|R|=1\\|S|=k_{2}\end{subarray}}||\sum \limits_{m\in R}|a_m\rangle \langle a_m|+\sum \limits_{n\in S}|b_n\rangle \langle b_n|||^2_{\infty})\\[2mm]
&+\max\limits_{\begin{subarray}{c}|R|=1\\|S|=k_{1}\end{subarray}}||\sum \limits_{m\in R}|a_m\rangle \langle a_m|+\sum \limits_{n\in S}|b_n\rangle \langle b_n|||^2_{\infty}].~~~~~~~~~~~~~~~~~~~~~~~~~~~~~~~~~~~~~~~~~~~~~~~~~~~~~~~~~~~~~~~~~~~~k=1,2,...,d. ~~~~~~~~~~~~~~~~~~~~~ \blacksquare
\end{aligned}
\end{equation}
\\

The solution given in Eq. (\ref{e:omega3}) can be explained as follows.
%in the above theorem, we would like to make a short description for the reader.
First, for $k=1, 2$, they are solved simply as
$$
\Omega_1=\frac{1}{4}[1+c]^2, \ \ \ \ \ \ \ \Omega_2=\frac{1}{4}[1+c^\prime]^2,
$$
where $c= \max \limits_{m,n}|\langle a_m|b_n\rangle|$, $c^\prime= \max \sqrt{|\langle a_m|b_n\rangle|^2+|\langle a_{m^\prime}{|b_{n^\prime}}\rangle|^2}$,
the maximum is taken over all indices $m\neq m^\prime$, $n\neq n^\prime$, and over all $n = n^\prime$, $m\neq m^\prime$. Then, for $k=3$ in JPDD,
$
\Omega_3 = \max \limits_{I_3} \max \limits_{\rho}\sum \limits_{(M,N)\in I_3} p_m(\rho)q_n(\rho)=\max\{\Omega_{3}^d,\Omega_{2,1}^d,\Omega^d_{1,1,1}\}.
$
Furthermore, $\Omega^d_3= \max \limits_{\begin{subarray}{c}|R|=1\\|S|=3\end{subarray}}\max \limits_{\rho} (\sum \limits_{m\in R}p_m(\rho))(\sum \limits_{n\in S}q_n(\rho))=\frac{1}{4}\max\limits_{\begin{subarray}{c}|R|=1\\|S|=3\end{subarray}}||\sum \limits_{m\in R}|a_m\rangle \langle a_m|+\sum \limits_{n\in S}|b_n\rangle \langle b_n|||^2_{\infty}$ and  $\Omega^d_{1,1,1}=\frac{1}{4}\max\limits_{\begin{subarray}{c}|R|=3\\|S|=1\end{subarray}}||\sum \limits_{m\in R}|a_m\rangle \langle a_m|+\sum \limits_{n\in S}|b_n\rangle \langle b_n|||^2_{\infty}.$ Since $\Omega_{2,1}^d=(\Omega_{1,1}^d-\Omega_1)+\Omega_{2}^d$,
we get
$
\Omega^d_2=\frac{1}{4}\max\limits_{\begin{subarray}{c}|R|=1\\|S|=2\end{subarray}}||\sum \limits_{m\in R}|a_m\rangle \langle a_m|+\sum \limits_{n\in S}|b_n\rangle \langle b_n|||^2_{\infty}
$
and
$
\Omega^d_{1,1}=\frac{1}{4}\max\limits_{\begin{subarray}{c}|R|=2\\|S|=1\end{subarray}}||\sum \limits_{m\in R}|a_m\rangle \langle a_m|+\sum \limits_{n\in S}|b_n\rangle \langle b_n|||^2_{\infty}.
$
We have shown how to calculate $\Omega_1$, $\Omega_2$ and $\Omega_3$. For the cases $k\geq 4$, interested readers can calculate $\Omega_k$ using a similar method and we sketch the details in the Methods.

The above theory enables us to formulate a series of $\Omega_k$, based on all quantities we obtain a tighter majorization vector $\omega$.
Note that our method is valid when all the maximums are taken over the same quantum state, otherwise our bounds will fail to hold.
Even so, our results can outperform $B_{Maj2}$ \cite{RPZ12} to some extent.
%Now, we remark that a very recent work, Pucha{\l}a, Rudnicki and \.{Z}yczkowski's work\cite{RPZ16}, has been completed independent on Ref. [15] to study uncertainty relation by majoration. More significantly, under the  circumstance of sum of two Shannon entropies in such work, the Maassen-Uffink bound \cite{Maassen10} has been improved in the whole range of its parameter and the improved result is stronger than Coles and Piani's result \cite{CP11}. Inspired by this reason, we compare our result under sum of two Shannon entropies with Pucha{\l}a, Rudnicki and \.{Z}yczkowski's work.

Our results enable us to strengthen the bounds on the sum of two Shannon entropies by $\mathbf{B_{JPDD}}=H(\omega)$, where $\omega$ is given by the improved $\Omega_k$ in Eq. (\ref{e:omega3}) and $H$ is the Shannon entropy. To see this phenomenon, let us first consider a $4$-dimensional system with incompatible observables $\{|a_m\rangle, |b_n\rangle\}_{m,n=1}^{4}$
\begin{equation*}
\scriptsize
\left\{
\begin{aligned}
(\sqrt{\frac{2}{3}}\cos \theta, \sqrt{\frac{2}{3}}\sin \theta, \sqrt{\frac{1}{3}}\cos \theta, \sqrt{\frac{1}{3}}\sin \theta), ~~(0.68, 0.72, 0.12, -0.071), ~~(-\sqrt{\frac{2}{3}}\sin \theta, \sqrt{\frac{2}{3}}\cos \theta, -\sqrt{\frac{1}{3}}\sin \theta, \sqrt{\frac{1}{3}}\cos \theta),~~(0.32, -0.47, 0.71, -0.42), \\ (-\sqrt{\frac{1}{3}}\cos \theta, -\sqrt{\frac{2}{3}}\sin \theta, -\sqrt{\frac{2}{3}}\cos \theta, \sqrt{\frac{2}{3}}\sin \theta),~~(-0.61, 0.48, 0.63, 0.064), ~~(\sqrt{\frac{1}{3}}\sin \theta, -\sqrt{\frac{1}{3}}\cos \theta, -\sqrt{\frac{2}{3}}\sin \theta, \sqrt{\frac{2}{3}}\cos \theta), ~~(-0.24, 0.19, -0.29, -0.9)
\end{aligned}
\right\},
\end{equation*}
and the unitary transformation $U$ ($U_{mn}=\langle a_m|b_n\rangle$) between them
\begin{equation*}
\left(
  \begin{array}{cccc}
    0.63 \cos \theta + 0.54 \sin \theta & 0.67 \cos \theta - 0.62 \sin \theta & -0.13 \cos \theta + 0.43 \sin \theta & -0.37 \cos \theta - 0.36 \sin \theta \\
    0.54 \cos \theta - 0.63 \sin \theta & -0.62 \cos \theta - 0.67 \sin \theta & 0.43 \cos \theta + 0.13 \sin \theta & 0.43 \cos \theta + 0.13 \sin \theta \\
    -0.30 \cos \theta - 0.47 \sin \theta & 0.4 \cos \theta - 0.072 \sin \theta & 0.86 \cos \theta - 0.23 \sin \theta & -0.098 \cos \theta - 0.85 \sin \theta \\
    -0.47 \cos \theta + 0.30 \sin \theta & -0.072 \cos \theta - 0.4 \sin \theta & -0.23 \cos \theta - 0.86 \sin \theta & -0.85 \cos \theta + 0.098 \sin \theta \\
  \end{array}
\right).
\end{equation*}
In Fig. 7, we plot the difference between $\mathbf{B_{JPDD}}$ and $B_{Maj2}$ \cite{RPZ12}, i.e. $B_{JPDD}-B_{Maj2}$ (the red line). Clearly, our bound $B_{JPDD}$ is tighter than $B_{Maj2}$ to some extent. Note that, the entropies are defined with base $2$ in general. But in our figures, in order to make them more readable, we take the natural logarithm instead.

To appreciate the stronger vector $\omega$ in the
improved UUR, we can consider Shannon entropy in the uncertainty relations to obtain a tighter bound than the previous work \cite{SR17}.% is not difficult.
Namely if we $\Phi$ in Eq. (\ref{phi}) as the Shannon entropy $H$, then we have

\begin{figure}[ht]
\centering
\includegraphics[width=0.5\textwidth]{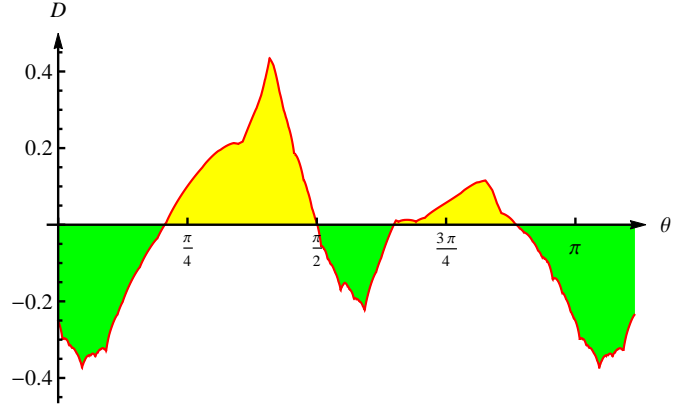}
\caption{{Difference between $B_{JPDD}$ and $B_{Maj2}$.}}
\end{figure}

\begin{eqnarray}\label{e:Shannon}
H(A)+H(B)\geq
\begin{cases}
-2\log c,       &\mathrm{if}~~~0<c\leq 1/ \sqrt{2}, \\
H_1(c),   &\mathrm{if}~~~1/ \sqrt{2}\leq c\leq c^{\ast}, \\
G(c),           &\mathrm{if}~~~c^{\ast}\leq c\leq 1,
\end{cases}
\end{eqnarray}
where $c^{\ast}\approx 0.834$, the bound $H_1(c)$ is the same as the one given in Vincente and Ruiz's work \cite{SR17},
and $G(c)=H(\omega)$ with $\omega$ given by our formula Eq. (\ref{e:omega1}). The bound $G(c)$ outperforms the Vincente and Ruiz's bound $F(c)$ \cite{SR17} in the interval $[c^{\ast}, 1]$. For further details, see Fig. 8.

\begin{figure}[ht]
\centering
\includegraphics[width=0.5\textwidth]{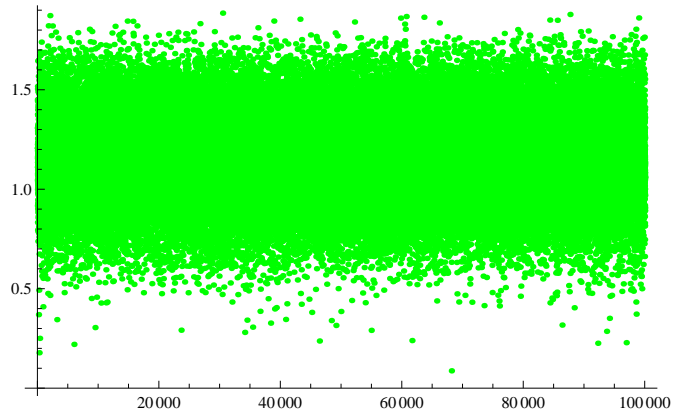}
\caption{{The vertical coordinate is $G(c)-F(c)$.
The horizontal coordinate is for random runs. It can be seen that
our bound outperforms the bound \cite{SR17} $100\%$ of the time,
while a bound given by  Friedland \cite{Friedland15} outperforms the bound \cite{SR17} around $90\%$ of the time.}}
\end{figure}

\section*{Conclusion}

In conclusion, we have presented a method called joint probability distribution diagram to strengthen the
bounds for the universal uncertainty relations. As an example, we consider the bounds on the sum of the Shannon entropies. As the universal uncertainty relations capture the essence of uncertainty in quantum theory, it is unnecessary to quantify them by particular measures of uncertainty such as Shannon or Renyi entropies.
Our results give a way to resolve some important cases in this direction, and is shown to offer a better bound for any uncertainty relations given by the nonnegative Schur-concave functions. Furthermore, how to extend this method to the case of multiple measurements are interesting, which requires further studies.

\section*{Methods}

{\bf Proof of the Lamma}~~~~~~The case of $k=1$ is obvious since the maximal element is $p_1q_1$.
Assume that the statement holds for the case of $k-1$: $\Omega_{k-1}=\Omega^d_{k_1,k_2,\cdots k_n}$ is connected with
$k_1>\cdots >k_n>0$. Suppose on the contrary that
the next maximum $\Omega_k=\Omega_{k-1}+p_iq_j$ is not connected.
Then there are two possibilities: (i) $i>n$ or $j>k_1$, thus we can replace $p_iq_j$ by
$p_iq_1$ or $p_1q_j$ and move further to $p_{n+1}q_1$ or $p_1q_{k_1+1}$ to get a possible bigger value
for $\Omega_k$. (ii) $i\leq n$ and $j>k_i$, in this case we can also replace the box $p_iq_j$ by
a connected box $p_{i'}q_{j'}$ to the region of $\Omega_{k-1}$ by sliding it leftward or upward.
Hence the statement is true by induction.~~~~~~~~~~~~~~~~~~~~~~~~~~~~~~~~~~~~~~~~~~~~~~~~~~
~~~~~~~~~~~~~~~~~~~~~~~~~~~~~~~~~~~~~~~~~~~~~~~~~~~~~~~~~~~~~~~~~~~~
~~~~~~~~~~~~~~~~~~~~~~~~~~~~~~~~~~~~~~~~~~~~~~~~~~~~~~~$\blacksquare$\\
~~\\

\noindent{\bf Proof of the Theorem}~~~~~~To calculate $\Omega_k=\Omega_{k_1,k_2,\cdots k_n}^d$, $k_1\geq \cdots \geq k_n$ and $k_1+\cdots+k_n=k$, we note that
\begin{align}\nonumber
\Omega^d_{k_1}&=\frac{1}{4}\max\limits_{\begin{subarray}{c}|R|=1\\|S|=k_1\end{subarray}}
||\sum \limits_{m\in R}|a_m\rangle \langle a_m|+\sum \limits_{n\in S}|b_n\rangle \langle b_n|||^2_{\infty},\\ \nonumber
\Omega^d_{k_2,k_2}&=\frac{1}{4}\max\limits_{\begin{subarray}{c}|R|=2\\|S|=k_2\end{subarray}}
||\sum \limits_{m\in R}|a_m\rangle \langle a_m|+\sum \limits_{n\in S}|b_n\rangle \langle b_n|||^2_{\infty}.
\end{align}
where $R$ and $S$ are subsets of distinct indices from [d], $|R|$ is the cardinality of $R$,
and $||\cdot||_{\infty}$ is the infinity operator norm which coincides with the maximum eigenvalue of the positive operator.
For a given $k$, there exist sets of $k_1\geq ...\geq k_n$ such that $\sum \limits_{i=1}^n k_i=k$ for some $n$.
For any such given $k_1\geq ...\geq k_n$, the quantity in $[.]$ in Eq. (\ref{e:omega3}) can be calculated.
The outer $\max$ picks up the largest quantity for all such possible $k_1,...,k_n$.
Then $\Omega_{k_1,k_2}^d=\Omega_{k_2,k_2}^d+(\Omega_{k_1}^d-\Omega_{k_2}^d)$
and $\Omega_{k_1,k_2,k_3}^d=\Omega_{k_3,k_3,k_3}^d+(\Omega_{k_1,k_2}^d-\Omega_{k_3,k_3}^d)$. Continuing in this way we have

\begin{align}\nonumber
\Omega_{k_1,\cdots, k_n}^d &=~\Omega^d_{\underbrace{k_n,\cdots,k_n}\limits_n}+(\Omega^d_{k_1,\cdots, k_{n-1}}-\Omega^d_{\underbrace{k_n,\cdots,k_n}\limits_{n-1}})\\ \nonumber
&=~\Omega^d_{\underbrace{k_n,\cdots,k_n}\limits_n}-\Omega^d_{\underbrace{k_n,\cdots,k_n}\limits_{n-1}}+
\Omega^d_{k_1,\cdots, k_{n-1}}\\ \nonumber
&=~(\Omega^d_{\underbrace{k_n,\cdots,k_n}\limits_n}-\Omega^d_{\underbrace{k_n,\cdots,k_n}\limits_{n-1}})+ (\Omega^d_{\underbrace{k_{n-1},\cdots,k_{n-1}}\limits_{n-1}}-\Omega^d_{\underbrace{k_{n-1},\cdots,k_{n-1}}\limits_{n-2}})\\ \nonumber
&~~~~+\cdots +(\Omega^d_{k_2,k_2}-\Omega^d_{k_2})+\Omega_{k_1}^d\\ \nonumber
&=~\frac{1}{4}[(\max\limits_{\begin{subarray}{c}|R|=n\\|S|=k_n\end{subarray}}||\sum \limits_{m\in R}|a_m\rangle \langle a_m|+\sum \limits_{n\in S}|b_n\rangle \langle b_n|||^2_{\infty}-\max\limits_{\begin{subarray}{c}|R|=n-1\\|S|=k_n\end{subarray}}||\sum \limits_{m\in R}|a_m\rangle \langle a_m|+\sum \limits_{n\in S}|b_n\rangle \langle b_n|||^2_{\infty})\\ \nonumber
&~~~~+
(\max\limits_{\begin{subarray}{c}|R|=n-1\\|S|=k_{n-1}\end{subarray}}||\sum \limits_{m\in R}|a_m\rangle \langle a_m|+\sum \limits_{n\in S}|b_n\rangle \langle b_n|||^2_{\infty}-\max\limits_{\begin{subarray}{c}|R|=n-2\\|S|=k_{n-1}\end{subarray}}||\sum \limits_{m\in R}|a_m\rangle \langle a_m|+\sum \limits_{n\in S}|b_n\rangle \langle b_n|||^2_{\infty})\\ \nonumber
&~~~~+
\cdots+(\max\limits_{\begin{subarray}{c}|R|=2\\|S|=k_{2}\end{subarray}}||\sum \limits_{m\in R}|a_m\rangle \langle a_m|+\sum \limits_{n\in S}|b_n\rangle \langle b_n|||^2_{\infty}-\max\limits_{\begin{subarray}{c}|R|=1\\|S|=k_{2}\end{subarray}}||\sum \limits_{m\in R}|a_m\rangle \langle a_m|+\sum \limits_{n\in S}|b_n\rangle \langle b_n|||^2_{\infty})\\  \nonumber
&~~~~+
\max\limits_{\begin{subarray}{c}|R|=1\\|S|=k_{1}\end{subarray}}||\sum \limits_{m\in R}|a_m\rangle \langle a_m|+\sum \limits_{n\in S}|b_n\rangle \langle b_n|||^2_{\infty}],
\end{align}

\noindent which gives the improved values of $\Omega_k=\max \Omega_{k_1,k_2,\cdots k_n}^d$, where max is taken over $\sum \limits_{i=1}^n k_i=k$ with $k_1\geq ...\geq k_n$,
and hence $\omega = (\Omega_1, \Omega_2-\Omega_1, \ldots, \Omega_d-\Omega_{d-1}, 0, \ldots, 0)$ for any $\rho$.
~~~~~~~~~~~~~~~~~~~~~~~
~~~~~~~~~~~~~~~~~~~~~~~~~~~~~~~~~~~~~~~~~~~~~~~~~~~~~~~~~~~~~~~~~~~~
~~~~~~~~~~~~~~~~~~~~~~~~~~~~~~~~~~~~~~~~~~~~~~~~~~~~~~~$\blacksquare$

\section*{Acknowledgements}

Supported by the Research Foundation for Youth Scholars of Beijing Technology and Business University QNJJ2017-03. The work is supported by NSFC (11275131, 11271138, 11305105, 11675113).

\section*{Author contributions statement}
T. Li, Y. Xiao, T. Ma, S.-M. Fei, N. Jing,  X.Q. Li-Jost and Z.-X wrote the main manuscript. All authors reviewed the manuscript.

\section*{Additional information}
Competing financial interests: The authors declare no competing financial interests.

\end{document}